\documentclass[prd, twocolumn, nofootinbib, floatfix]{revtex4-1}

\usepackage{amsmath}
\usepackage{graphicx}
\usepackage{dcolumn}
\usepackage{bm}
\usepackage{enumerate} 
\usepackage{epsfig}
\usepackage{amssymb,latexsym,mathrsfs}
\usepackage{graphicx}
\usepackage{color}
\usepackage{hyperref}

\hypersetup{
    colorlinks=true,
    linkcolor=red,
    citecolor=blue,
} 

\newcommand{\be}{\begin{equation}}
\newcommand{\ee}{\end{equation}}
\newcommand{\bes}{\begin{equation*}}
\newcommand{\ees}{\end{equation*}}
\newcommand{\beq}{\begin{equation}}
\newcommand{\eeq}{\end{equation}}
\newcommand{\bs}{\begin{split}} 
\newcommand{\bea}{\begin{eqnarray}}
\newcommand{\eea}{\end{eqnarray}}
\newcommand{\beqa}{\begin{eqnarray}}
\newcommand{\eeqa}{\end{eqnarray}}

\newcommand{\om}{\Omega_m}
\newcommand{\Om}{\Omega_m}

\newcommand{\lcdm}{$\Lambda$CDM} 
\newcommand{\alens}{A_{\rm lens}}

\begin{document}

\title{CMB Lensing and Scale Dependent New Physics} 
\author{Alireza Hojjati$^{1,2}$, Eric V.\ Linder$^3$} 
\affiliation{$^1$Department of Physics and Astronomy, University of 
British Columbia, Vancouver, V6T 1Z1, BC, Canada \\
$^2$Physics Department, Simon Fraser University, Burnaby, V5A 1S5, BC, Canada\\
$^3$Berkeley Center for Cosmological Physics \& Berkeley Lab, 
University of California, Berkeley, CA 94720, USA} 

\begin{abstract}
Cosmic microwave background lensing has become a new cosmological 
probe, carrying rich information on the matter power spectrum and 
distances over the redshift range $z\approx1$--4. We investigate the 
role of scale dependent new physics, 
such as from modified gravity, neutrino mass, and cold (low sound speed) 
dark energy, and its signature on CMB lensing. The distinction between 
different scale dependences, and the different redshift dependent 
weighting of the matter power spectrum entering into CMB lensing and 
other power spectra, imply that CMB lensing can probe simultaneously 
a diverse range of physics. We highlight the role of arcminute resolution 
polarization experiments for distinguishing between physical effects. 
\end{abstract}

\date{\today} 

\maketitle

\section{Introduction} 

Gravitational lensing of the cosmic microwave background radiation (CMB 
lensing) is a recently measured, powerful cosmological 
probe. The primordial photons are deflected by mass concentrations along the 
line of sight, sampling the matter density power spectrum -- and the laws 
of gravity -- over the entire cosmic history from recombination to the 
present. While this rearrangement of photons has long been recognized, 
with early papers accounting for the key elements of both the dispersive and 
coherent nature of 
the scattering, and its dependence on the matter power spectrum, dating to 
the 1980's \cite{lin88,coleefst,sasaki,linsky}, the first statistically 
significant detection in the CMB alone was in 
2011 \cite{das2011}. 

The lensing smears out the photon temperature power spectrum, but also 
induces non-Gaussianity, generating nontrivial four-point correlations 
\cite{fourpt1,fourpt2}, and a form of parity violation, converting between 
$E$-mode (parity even) and $B$-mode (parity odd) polarization \cite{polzn}. 
These effects have now all been detected 
\cite{das2011,plancklens13,plancklens15,act2,spt1,spt2,spt3,pb1,pb2,pb3,bkp}. 

From these observed effects one forms the CMB lensing power spectrum, a 
measure of the lensing strength as a function of the multipole, or angular 
scale. This will be the tool we focus on in this paper to explore new 
physics. One can compare this to, in the first instance, complete lack of 
lensing (i.e.\ verifying that lensing exists), and then for example, to the 
power predicted in the $\Lambda$CDM model as a test of the cosmology. Current 
constraints on the CMB lensing power spectrum include 
\cite{plancklens15,spt14124760,pb2,act13011037,spt1}.

With the first measurements, the statistical significance of detection 
relative to a null result was the main result. This was often quoted in 
terms of the ratio of the measured lensing power relative to that predicted 
in the concordance cosmological constant plus cold dark matter (\lcdm) 
cosmology fit from the temperature power spectrum: 
$\alens=C^{dd}/C^{dd}_{\Lambda{\rm CDM}}$ 
\cite{alensgfs}. Now that measurements of the lensing power spectrum have 
dramatically improved, to signal to noise levels greater than 40, the 
characterization of the detailed power spectrum is of interest. 

While one might consider the amplitude of the lensing power as a measure 
of the overall growth of matter clustering, this is not quite true: the 
lensing power spectrum is a projection from many redshifts, hence the growth 
rate effectively enters, and over many wavenumbers to a given angular 
multipole $\ell$, 
so that nonlinear density evolution can enter even at low, ostensibly 
linear $\ell$. Thus any change in cosmology should exhibit a different 
angular dependence than in the given concordance model, at some level. This 
implies that $\alens$ becomes scale dependent. 

Moreover, when scale dependent growth arises even in the linear density 
regime, we expect a correspondingly 
stronger signature of scale dependence (relative to 
\lcdm) in the CMB lensing power spectrum. Thus, CMB lensing can act as a 
probe of such physics, i.e.\ modified gravity with its scalaron Compton 
wavelength, neutrino mass with its free streaming scale, or clustering dark 
energy with its sound horizon. 

Indeed, \cite{plancklens15} recently measured $\alens$ in several 
bins of $\ell$ (see their Table~1 and Fig.~6; also see Fig.~34 of 
\cite{1507.02704}) with mild hints of scale dependence. Increased 
accuracy, especially 
from ongoing high resolution, ground based polarization experiments such 
as the Atacama Cosmology Telescope \cite{act}, POLARBEAR/Simons Array 
\cite{polarbear}, and the South Pole Telescope \cite{spt}, and the next 
generation CMB-S4, will place 
constraints on such scale dependent effects. 

In Sec.~\ref{sec:linear} we review the basic relation of the CMB lensing 
power spectrum to the matter power spectrum and gravitational coupling 
strength. We present a simple analytic expression to motivate intuition 
for the expected scale dependent $\alens$ in the ideal, large scale linear 
limit in Sec.~\ref{sec:anly}. 
For full numerical results we adapt the Boltzmann code MGCAMB in 
Sec.~\ref{sec:code} and investigate the lensing power spectrum for several 
sources of scale dependent physics. We conclude in Sec.~\ref{sec:concl}.

\section{CMB lensing power spectrum} \label{sec:linear} 

In this section we give a brief review of CMB lensing and its relation to 
the matter power spectrum and the gravitational coupling strength. We also 
illustrate the role of the projection of the matter power from different 
redshifts and different wavenumbers onto the lensing deflection power 
spectrum observed at a given angular multipole. 

As in weak gravitational lensing of background sources such as galaxies, 
the angular power spectrum of the lensing potential $\phi$ is given by 
(see, e.g., \cite{0601594}) 
\be 
C_\ell^{\phi\phi}=\frac{8\pi^2}{\ell^3}\int_0^{\chi_{\rm lss}} d\chi\, 
\chi \left(\frac{\chi_{\rm lss}-\chi}{\chi\chi_{\rm lss}}\right)^2 
P_{\Psi+\Phi}\left(k=\frac{\ell}{\chi};\chi\right) \ . 
\ee 
Here $\chi$ is the comoving distance to the lens, $\chi_{\rm lss}$ is the 
comoving distance to the CMB last scattering surface, for simplicity we 
assume a spatially flat universe, $k$ is the Fourier wavenumber, 
and $P_{\Psi+\Phi}$ 
is the power spectrum of the sum of the time-time and space-space metric 
gravitational potentials $\Psi$ and $\Phi$. Thus lensing explicitly depends 
on cosmic geometry and gravity as well as growth. 
Note that unlike galaxy lensing, we do not 
have to integrate over the (inexactly known) source distribution since 
for the CMB the source is the well defined last scattering surface. This is 
an advantage of CMB lensing, in addition to its precision measurements. 

Since the lensing potential is not directly observable, we use the 
deflection vector ${\mathbf d}=\nabla\phi$. This is also related to 
the convergence $\kappa=-(1/2)\nabla\cdot{\mathbf d}=-(1/2)\nabla^2\phi$. 
In Fourier space, the power spectra will be related by 
$C^{\kappa\kappa}_\ell=\ell(\ell+1)C^{dd}_\ell=[\ell(\ell+1)]^2 C^{\phi\phi}_\ell$. 

The potential power spectrum can be related to the matter density power 
spectrum through 
\be 
P_{\Psi+\Phi}(k,z)=\frac{9\om^2(z) H^4(z)}{8\pi^2}\, 
\frac{G_{\rm eff}^{\Psi+\Phi}(k,z)}{G_N}\,k^{-1} P_\delta(k,z) \ , 
\label{eq:ppsi} 
\ee 
where $\om(z)$ is the dimensionless matter density, $H$ is the Hubble 
parameter, $z$ is the redshift, 
and $G_{\rm eff}$ reflects 
that in modified gravity the gravitational strength may not be Newton's $G_N$, 
modifying the Poisson equation. 
A convenient final expression for the convergence power spectrum is \cite{pb1} 
\be 
C_\ell^{\kappa\kappa}=\int dz\,\frac{H(z)}{\chi^2} W^2(z)\,P_\delta(k=\ell/\chi) 
\ . \label{eq:clpk} 
\ee 
Here $W$ is a window function, or kernel. It has a broad peak roughly halfway 
to the last scattering surface, and so CMB lensing has substantial sensitivity 
from $z\approx0.5$--5, allowing it to probe matter and gravity to higher 
redshifts than many other observables. 

(As an aside, note that converting $\om(a)H^2(z)$ 
to $\om H_0^2 a^{-3}$ in Eq.~\ref{eq:ppsi} and pulling the present matter 
density $\om$ outside 
the integral in Eq.~\ref{eq:clpk} is not valid in models that introduce 
a matter coupling \cite{11025090}.) 

\begin{figure}[htbp!]
\includegraphics[width=\columnwidth]{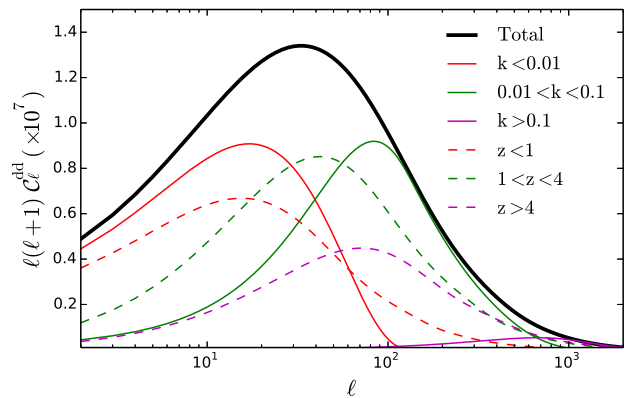} 
\caption{
The lensing deflection power spectrum for a concordance \lcdm\ 
cosmology with $\om=0.3$ (main, black curve) 
is plotted vs multipole, showing a broad peak 
around the coherence scale of $\sim2^\circ$, but with power over a range of 
scattering angles down to the typical deflection of a few arcminutes. 
The shorter solid curves within the main envelope illustrate the 
contributions of different Fourier modes $k$ to the deflection spectrum; 
the dashed curves show the contributions of different redshift windows. 
} 
\label{fig:cell} 
\end{figure}

Figure~\ref{fig:cell} shows the CMB lensing deflection power spectrum for 
a concordance \lcdm\ cosmology, exhibiting the main characteristics 
(also see the pioneering Figs.~3 and 4 of \cite{0601594}). 
While the peak is at $\ell\approx40$, it 
extends over a broad range of multipoles. Because of the projection in both 
redshift and wavenumber, a given $\ell$ does not correspond to a unique 
length scale in the matter power spectrum, or a unique time in the growth 
of density perturbations. This is important in its effect of blending linear 
and nonlinear physics. 

We indicate what portion of the deflection spectrum 
arises from lensing in different redshift ranges, and also different Fourier 
wavenumbers. The general rule of thumb is that small $k$ (large scales) 
corresponds to low $\ell$, and low redshift corresponds to large angles for 
a given scale and hence also low $\ell$. However, while high $k$ modes that 
are beyond the linear regime will dominate the high $\ell$ spectrum, because 
of projection ($\ell=k\chi$) they can also influence the lower $\ell$ region. 
Thus nonlinear, and scale dependent, physics can leave its mark over a wide 
range of the deflection power spectrum. 

\begin{figure}[htbp!]
\includegraphics[width=\columnwidth]{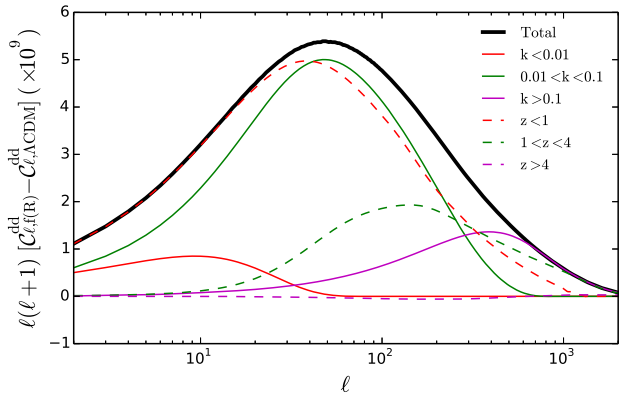} 
\caption{
The deviations in the lensing deflection power spectrum between a \lcdm\ 
and $f(R)$ gravity cosmology are shown for the total power and for various 
windows in redshift and wavenumber. Scale dependence can change the 
redshift and Fourier mode weighting (here exaggerated by taking 
$B_0=0.01$, see Sec.~\ref{sec:code} for details). 
} 
\label{fig:cellb0} 
\end{figure}

Figure~\ref{fig:cellb0} gives one example of the difference that scale 
dependent physics can make to the redshift and wavenumber weighting. 
This shows the deviation from the \lcdm\ case for $f(R)$ scalar-tensor 
gravity, as discussed in Sec.~\ref{sec:code}. 
In this particular case, the contribution from $z<1$ is most 
strongly affected since the modified gravity restores to general relativity 
at high redshift. In wavenumber, there are different modifications in 
the low, quasilinear, and high Fourier mode regimes. We give further 
examples, for different physical origins for scale dependence, in the 
next section.

\section{Analytic scale dependence} \label{sec:anly} 

In the linear density perturbation regime, the matter power spectrum 
in the standard model 
evolves with a scale independent growth factor, only changing its amplitude. 
The redshift dependent projection of different Fourier modes $k$ onto 
multipoles $\ell$, however, means that a scale dependence in the lensing 
power spectrum is induced nevertheless from any change in the growth that 
arises from a change in the cosmic expansion. (Alteration of growth due purely 
to a uniform multiplication of the gravitational strength can indeed give a 
scale independent $\alens$ -- indeed this is what \cite{alensgfs} 
originally considered.)

\begin{figure}[htbp!]
\includegraphics[width=\columnwidth]{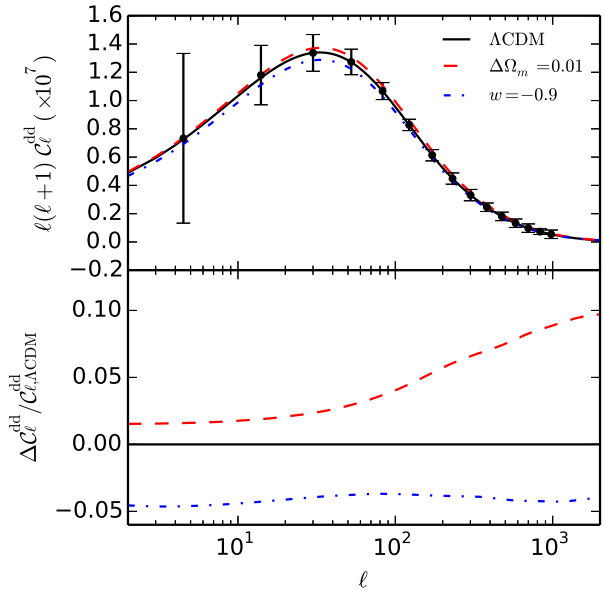} 
\caption{The effects of changes in the matter density and dark energy 
constant equation of state on the lensing deflection power spectrum are 
shown. Since the influence of these parameters on the linear growth factor 
is scale independent, the resulting deviations in the deflection power 
spectrum are nearly scale independent -- at low $\ell$ where linear modes 
are untainted by nonlinear modes. The error bars show uncertainties 
from Planck 2015 results \cite{plancklens15}, centered on the \lcdm\ theory 
curve. The current constraints have uncertainties greater than 5-10\% per 
bandpower and cannot distinguish smaller effects. 
} 
\label{fig:dom} 
\end{figure}

Figure~\ref{fig:dom} demonstrates this scale dependence. It is generally 
quite mild, which is why $\alens$ was initially a reasonable parametrization 
at the signal to noise levels first obtained. We see that the scale 
dependence is a few percent 
effect out to $\ell\sim100$, past the peak of the deflection power spectrum, 
for a change $\Delta\om=0.01$ or $\Delta w=0.1$, 
where $w$ is the constant dark energy equation of state parameter. 
It was only very recently that measurement uncertainties dipped below 
the 10\% level.

Several types of cosmological physics, however, can induce scale dependence 
in the matter growth even in the linear regime. Examples include modified 
gravity, such as scalar-tensor theories, which have a Compton wavelength 
associated with them, neutrino mass, which has a free streaming scale, and 
cold (and hence clustered) dark energy, which has a sound horizon. In this 
section we present a simple analytic treatment purely to build our intuition 
for the numerical results of the next section. 

Note that \cite{14065459} presents a careful analysis of the effect of 
the physical matter density on the CMB lensing deflection spectrum. 
Their fitting formula contains an implicit scale dependence through the 
local slope of the deflection spectrum. 
Explicit scale dependent 
physics from neutrino mass has recently been considered in \cite{150607493}. 

As we will see, our results for these cases are in good agreement with 
theirs, especially taking into account different treatments of the nonlinear 
regime (they use purely linear modes) and of other parameters (we fix all 
parameters except the one we are plotting, for clarity in exhibiting its 
effect; but see Sec.~\ref{sec:code} regarding preserving the acoustic scale 
instead). 

As our first example of scale dependent physics, consider scalar-tensor 
gravity. Here the Poisson equation relating the 
lensing potential $\Psi+\Phi$ to the matter density perturbation is 
unaffected, but the equation governing the growth of the density is changed. 
Theories like $f(R)$ gravity involve a particular scale dependence, and 
the gravitational coupling in the density growth equation can be written as 
\be 
\frac{G_{\rm eff}^\Psi}{G_N}=1+\frac{1}{3}\frac{1}{1+[aM(a)/k]^2} \ , 
\label{eq:geff} 
\ee 
where $M$ is the mass of the scalaron and $1/M$ its Compton wavelength. 
On scales larger than the Compton wavelength, the coupling is restored to 
Newton's constant, or equivalently at early times when the scalaron mass 
is large then the theory approaches general relativity. On smaller scales, 
however, gravity is strengthened (while on much smaller scales a chameleon 
screening mechanism can enter, again restoring to general relativity). 

Such changes to the source term of the growth equation can be treated in 
the formalism of \cite{lincahn} to determine the influence 
on growth to lowest order. From Eq.~(21) of \cite{lincahn}, where their 
$Q(k,a)={G_{\rm eff}^\Psi}/{G_N}$, 
we have 
\bea 
P_\delta(k)&=&P_{\delta,{\rm GR}}(k)\, 
\left[1+\int_0^a \frac{da'}{a'}\,(a'^4 H)^{-1} \right. \notag\\ 
&\qquad& \left. \int_0^{a'} \frac{da''}{a''}\,\frac{a''^4 H\Om(a'')}{1+[a''M(a'')/k]^2}\right]^2 \\ 
&\approx& P_{\delta,{\rm GR}}(k)\,
\left[1+k^2 \int_0^a \frac{da'}{a'}\,(a'^4 H)^{-1} \right. \notag\\ 
&\qquad& \left. \int_0^{a'} \frac{da''}{a''}\,\frac{a''^2 H\Om(a'')}{M^2}\right] \\
&\approx&P_{\delta,{\rm GR}}\,[1+k^2\,p(a)] \ . 
\label{eq:pk}
\eea 

The last two lines with the approximate signs keep only the lowest order 
terms in $k$, with the last line illustrating the leading order $k^2$ 
dependence of the modification. 
(Sound horizon terms also enter as $k^2$, while neutrino free streaming 
gives a different dependence but the general formalism still applies.) 
From Eq.~(\ref{eq:clpk}) we note that 
an additional $k^2$ term 
in $P_\delta(k)$ does not simply create an additional $\ell^2$ modification of 
the lensing power spectrum. Instead it reweights $C_\ell$ from the mass power 
spectrum at each redshift. That is, 
\bea 
\label{eq:cldd_ratio}
\frac{C^{dd}_\ell}{\bar C^{dd}_\ell}&=&1+\frac{\ell^2}{\bar C^{dd}_\ell} 
\int dz\,\frac{H(z)}{\chi^2} W^2(z)\,\bar P_\delta\,\frac{p(z)}{\chi^2}\notag 
\\ 
&\approx&1+\left\langle\frac{p(z)}{\chi^2}\right\rangle \, 
\eea 
where a bar indicates the unmodified (GR) case and angle brackets indicate 
a weighting over redshift, accounting for $k=\ell/\chi$. 
This is an important point, and means that scale dependent physics does 
not necessarily have an obvious form. 

Where might we expect the largest modification in the deflection power 
spectrum? This will be addressed numerically 
in the next section but here we can gain some intuition. Modified gravity 
that is consistent with other observations becomes important only fairly 
recently. For example, from Fig.~6 of \cite{linfr} we see that a modification 
in the matter spectrum by 10\% today may have been less than 1\% at $z=1$. 
Equivalently, Fig.~5 of \cite{linfr} shows that the parameter $B$ related to 
the Compton wavelength can easily be one to two orders of magnitude smaller 
at $z=1$ than its value $B_0$ today. As discussed previously, low redshift 
lensing contributed most to low multipoles, but so do low $k$ modes where 
the modification is suppressed by $k^2$. Conversely, higher $k$ modes where 
modified gravity effects are more important should appear at high $\ell$, 
but here one also has high redshift lensing contributions where modified 
gravity is diminished. Thus, the modified contributions are diluted by the 
unmodified ones. Since there are many more high $k$ modes than low $k$ ones, 
one might expect that the modifications do grow with $\ell$, but much more 
slowly than a naive $k^2\to\ell^2$ scaling, and that a low $\ell$ tail 
should be present. At high enough $\ell$ (roughly Mpc scales, and low 
redshift, so $\ell\gtrsim10^3$), the screening mechanism should enter and 
the deflection power spectrum approach that of \lcdm. 
The numerical computations bear this out. 

Let us further our intuition for the results by briefly considering 
neutrino mass, cold (clustered) dark energy, and standard matter density 
nonlinearities. If we compare models with different 
sums of the neutrino masses, we have to specify how we are compensating for 
this energy density in order to retain a total dimensionless energy density 
of unity, i.e.\ a spatially flat universe. If we trade matter density ($\Omega_m h^2$) for 
neutrino mass, then on large scales the nonrelativistic neutrinos act in the 
same manner as the subtracted matter, and we expect no significant effect. 
However on small scales, where the neutrinos free stream, we not only erase 
gravitational potentials from the free streaming but also reduce the matter 
clustering since there is less matter; these effects together should reinforce 
each other to cause substantial suppression of the deflection power spectrum. 
If the extra neutrino mass is compensated by reduced dark energy density 
($\Omega_\Lambda h^2$), then 
the neutrino free streaming is replacing what would anyway have been 
suppression due to the dark energy negative pressure, and so the effect should 
be more mild.  

\begin{figure}[htbp!]
\includegraphics[width=\columnwidth]{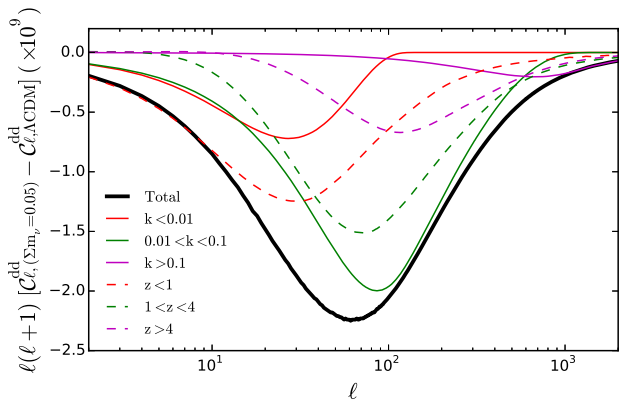}\\ 
\caption{
The deviations in the lensing deflection power spectrum between \lcdm\ 
and a model where part of the CDM energy density is replaced with that of massive neutrinos with $\sum m_\nu=0.05\,$eV are shown for the total power and for various 
windows in redshift and wavenumber.  The CMB lensing power is suppressed due 
to neutrino free streaming and less CDM clustering.  A similar trend exists for the case where part of the dark energy density is replaced with that of massive neutrinos.} 
\label{fig:cellmnu} 
\end{figure}

Figures~\ref{fig:cellmnu} illustrate how the lensing deflection power 
spectrum changes due to the presence of massive neutrinos for the case
of matter compensated neutrino mass. The results are fairly similar in the case of dark energy compensated neutrino mass except that, as stated above, the matter compensated case has 
a stronger effect. Most of the change in power due to neutrino mass 
is on intermediate scales near the free streaming scale. This, 
together with the projection between $k$ and $\ell$, explains the pattern 
seen in the figure.

As for dark energy, it can only clump on scales between the sound horizon 
and the Hubble scale, so a low sound speed (cold dark energy) is necessary 
for its clumping (as well as an 
equation of state significantly different from $w=-1$ at some epoch). If the 
dark energy clusters it can add to the deflection power, and furthermore the 
value of $w\ne-1$ removes some of the suppression of power, so cold 
(clustered) dark energy could leave an enhancement signature in the 
deflection power spectrum at low $\ell$. 

\begin{figure}[htbp!]
\includegraphics[width=\columnwidth]{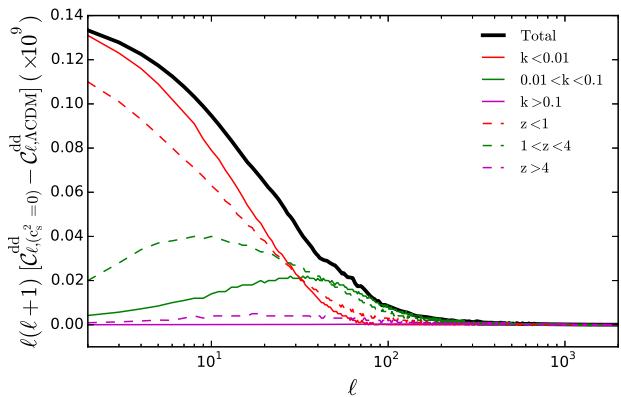} 
\caption{
The deviations in the lensing deflection power spectrum between \lcdm\ 
and a cold dark energy model are shown for the total power and for various 
windows in redshift and wavenumber.  In this case the CMB lensing power is 
enhanced, mostly on large scales and at late times. 
} 
\label{fig:celllowDE} 
\end{figure}

Figure~\ref{fig:celllowDE} summarizes these effects. It can be seen that 
the influence is predominantly on large scales ($k < 0.1\,h$/Mpc), where 
the additional clustering enhances the deflection power spectrum. Moreover 
this is mostly relevant at low redshift ($z<1$) when dark energy dominates 
the energy budget of the Universe. 

Finally, density perturbation growth beyond the linear regime exhibits scale 
dependence. This enters at $k\gtrsim0.1\,h/$Mpc in the matter power spectrum, 
which normally we would translate to roughly $\ell\gtrsim1000$ 
-- however, we expect that due to projection effects even lower $\ell$ can 
show noticeable effects. (See Fig.~\ref{fig:cell}, where 
the total power begins to diverge from the $k<0.1$ power around 
$\ell\approx200$, even sooner for higher $\om$.) 
In addition, nonlinearity is more prevalent at low redshifts, also pushing 
the influence to lower $\ell$. 
We will address this numerically in the next section.

\section{Numerical scale dependence} \label{sec:code} 

To explore all these effects we carry out a full numerical computation 
using the Boltzmann code MGCAMB (see \cite{mgcamb,Hojjati:2012ci} for 
details). MGCAMB is a modified version of the CAMB code \cite{camb} which 
includes a general parametrization of modified gravity theories in the 
linear regime of perturbations. This will enable us to investigate the 
scale dependence of $C^{dd}_\ell$ in the cases of modified gravity theories, 
neutrino mass, and cold (low sound speed) dark energy.

As a specific modified gravity theory we choose $f(R)$ models with the 
action
\begin{equation}
S = \frac{1}{16 \pi G}\int d^4x \sqrt{-g}\,\left[R+f(R)+\mathcal{L}_{\rm m}\right] \ .
\label{fRaction}
\end{equation}
These models can be tuned to reproduce any background expansion history, and the remaining relevant quantity is the squared Compton wavelength of the new scalar degree of freedom $f_R \equiv df /dR$ mediating the fifth force. In units of the Hubble length squared it is given by \cite{Song:2006ej,Hu:2007nk} 
\begin{eqnarray}
B \equiv {f_{RR} \over 1+f_R} {d R \over d\ln a} \left(  d\ln H \over d\ln a \right)^{-1}\,.
\end{eqnarray}
For a fixed background expansion history, different $f(R)$ models can be 
parametrized by the parameter $B_0$, which is related to the present value of 
the scalaron parameter $f_{R0}\approx -B_0/5$. The exact relation depends on 
the specific $f(R)$ model. 

Note that because of the rapid evolution of the curvature, a small value 
$B(z)$ could correspond to a much larger $B_0$ or $f_{R0}$. For example, 
in the exponential gravity model \cite{linfr} 
$B_0=0.01$ could come from $B(z=1)=10^{-4}$, 
and hence show little growth modification at high redshift. 
Regarding the expansion history, the maximum deviation of the 
dark energy equation of state $|1+w_{\rm max}|\approx B_0/2$ for exponential 
gravity, essentially giving \lcdm\ behavior. 
For examples of these relations, see Fig.~5 of \cite{linfr}. Thus, not 
all modified gravity theories will give clear signatures in early growth or 
in expansion. We consider instead a commonly used model, essentially the 
easiest case to constrain: Hu-Sawicki $f(R)$ gravity with $n=1$ 
\cite{Hu:2007nk}. 
As $n$ gets larger, $B$ evolves more rapidly, similar to the exponential 
gravity case, and becomes harder to constrain. 

In evaluating modified gravity effects it is important to use a robust 
code and not assume aspects of standard \lcdm\ growth.  MGCAMB has 
been tested against an independent Boltzmann code EFTCAMB \cite{eftcamb}, 
for example in \cite{14051022}, and the results are in 
good agreement. While MGCAMB evaluates the evolution equations in the 
quasistatic limit (while EFTCAMB treats them generally), this is an 
excellent approximation on the scales of interest to us 
\cite{12106880,13125309,14116128}. The lensing analysis in \cite{150201590} 
also confirms the quasistatic approximation is good. MGCAMB can also, in 
principle, include the nonlinear matter power spectrum unlike the 
intrinsically linear EFTCAMB formalism. We have seen that nonlinearities 
enter already at $\ell\approx200$. To incorporate the nonlinear density 
behavior for the $f(R)$ model, we employ the MGhalofit patch 
\cite{mghalofit}, which has been calibrated from simulations to correctly 
account for the modified gravity effects. 

\begin{figure*}[htbp!]
\includegraphics[width=0.8\textwidth]{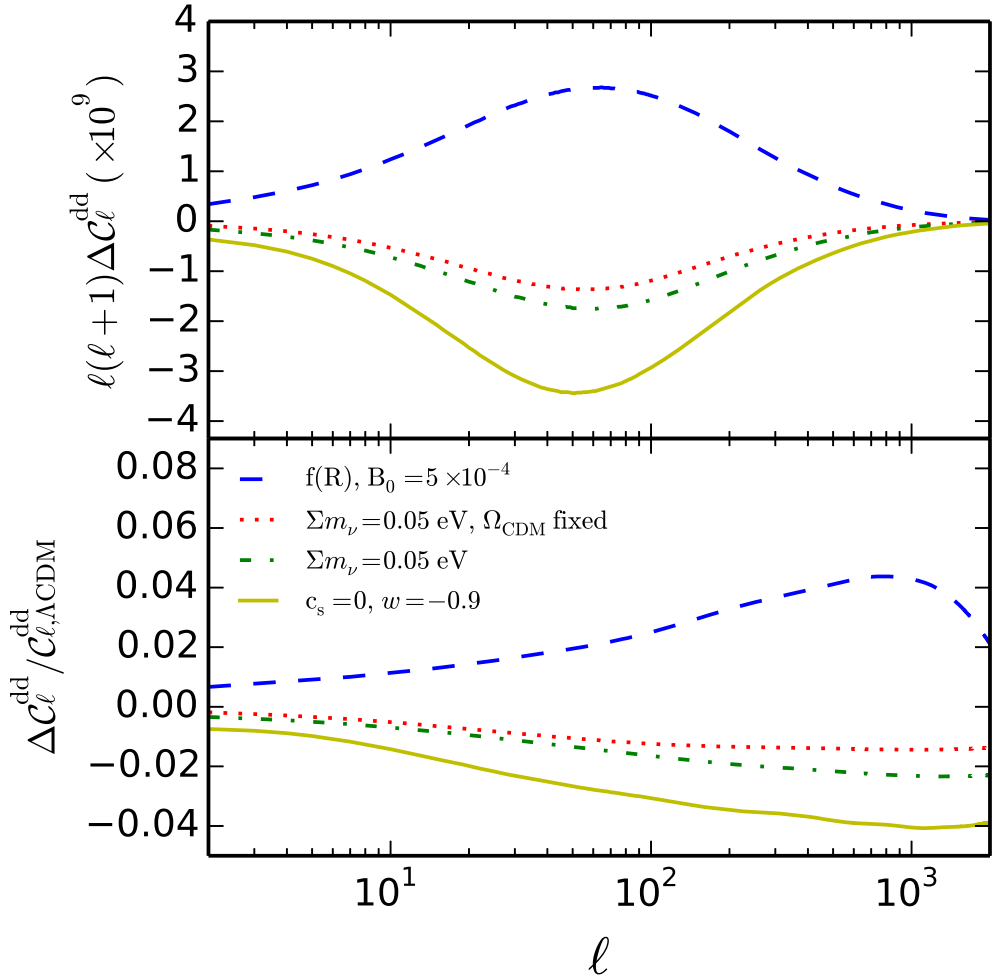} 
\caption{The CMB lensing deflection power spectrum is plotted vs multipole 
$\ell$ for several models with scale dependent physics, including modified 
gravity with a Compton scale $B_0$, neutrinos with a sum of masses 
$\sum m_\nu$, and cold dark energy with a sound speed $c_s$. 
} 
\label{fig:ratios} 
\end{figure*}

Fig.~\ref{fig:ratios} compares the CMB lensing power spectra of the 
scale dependent models discussed in the previous section to that from the 
fiducial ($\Lambda$CDM) model. 
In the $f(R)$ modified gravity model, we see an enhancement of power 
(as expected since scalar-tensor theories strengthen the gravitational 
coupling) on all scales. At low multipoles the effect is weaker, since 
on large scales, greater than the Compton wavelength, the coupling 
approaches Newton's constant (see Eq.~\ref{eq:geff}). At high multipoles 
(which recall include very nonlinear scales), the chameleon screening 
enters and again the result goes toward the general relativity case. 
Thus the expectation of Sec.~\ref{sec:anly} is borne out: the power 
approaches the \lcdm\ value on scales above the scalaron wavelength and 
below the screening scale. 
However in general we see a modification of CMB lensing power over a wide 
range of $\ell$ from a range of redshifts.

For the impact of massive neutrinos, recall we distinguished two cases. 
When we add massive neutrinos and keep $\om h^2$ fixed (retaining spatial 
flatness by decreasing $\Lambda$), we see that free streaming suppresses 
lensing power at small scales, smaller than the free streaming scale, as 
expected. However, when the added energy density of the massive neutrinos 
is compensated by decreasing that of the CDM instead, the effect is more 
pronounced on all scales. Not only is there suppression from free streaming 
but there is also less clustering due to the lower $\om$. 

\begin{figure}[htbp!]
\includegraphics[width=\columnwidth]{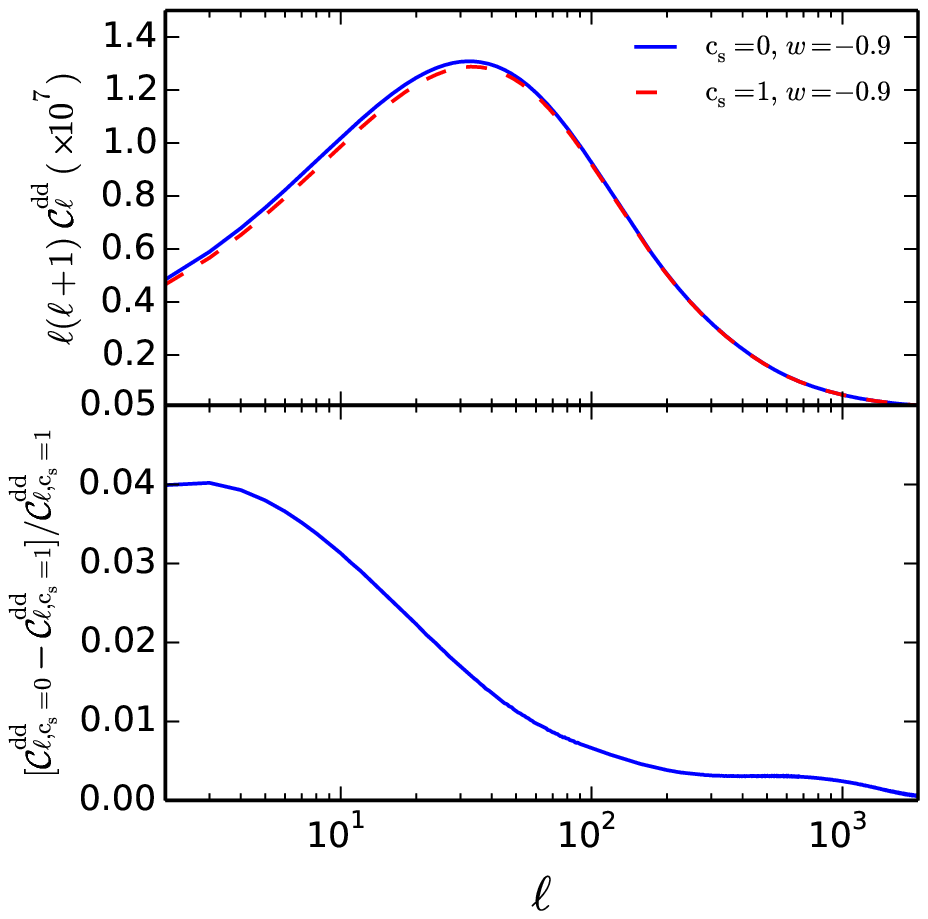} 
\caption{Cold dark energy provides extra clustering, and hence deflection 
power, on scales larger than the sound horizon (hence low multipoles). 
This also leads to scale dependence in the ratio of the power to that 
of \lcdm, or quintessence. The bottom panel shows the deviation of cold 
dark energy from quintessence with the same value $w=-0.9$, but $c_s=1$. 
} 
\label{fig:cde} 
\end{figure}

Scale dependence can also arise due to density perturbations above the sound 
horizon of cold dark energy with a low sound speed. As predicted in 
Sec.~\ref{sec:anly} the effects of low sound speed enter at low multipoles. 
At higher multipoles, \cite{09081585} showed that one could rescale the CMB 
lensing deflection spectrum by a uniform factor $\alens$ to a good 
approximation to take into account the deviation of $w$ from $\Lambda$. 
This agrees well with Fig.~\ref{fig:dom}, where we see that the deviation 
in power due to $w\ne-1$ is almost perfectly scale independent. 
If in Fig.~\ref{fig:ratios} one adjusted the high multipoles to match \lcdm, 
then the low multipoles show the expected power gain due to the clustering 
of cold dark energy. We make this more explicit in Fig.~\ref{fig:cde}.

Note that the shapes of the deviations in the CMB lensing spectrum are 
distinct between the different scale dependent physics. Modified gravity 
gives a rise and fall, while neutrino mass leads to an almost tanh like 
behavior of a low $\ell$ plateau, then nearly linear slope, then a high 
$\ell$ plateau, and cold dark energy gives a slightly more curved version 
of this, and one that is not scale independent at high $\ell$. 
(We have checked that a change in the Hubble constant to 
preserve the CMB acoustic scale does not appreciably change the shapes 
though it does increase the amplitude of the deviations.) 

The differing scale dependences would allow for identification of the 
non-\lcdm\ physics with sufficiently precise measurements. Here we have 
concentrated solely on the deflection power spectrum. These effects of 
course also show up in the matter power spectrum (which causes the 
deflections) but there the measurement through galaxy surveys has to 
contend with the 
expected scale dependent galaxy bias. CMB lensing also affects other 
CMB power spectra, giving secondary contributions to them involving the 
extra physics, though these will be mixed with the (unlensed) 
primordial perturbations. 

\begin{figure}[htbp!]
\includegraphics[width=\columnwidth]{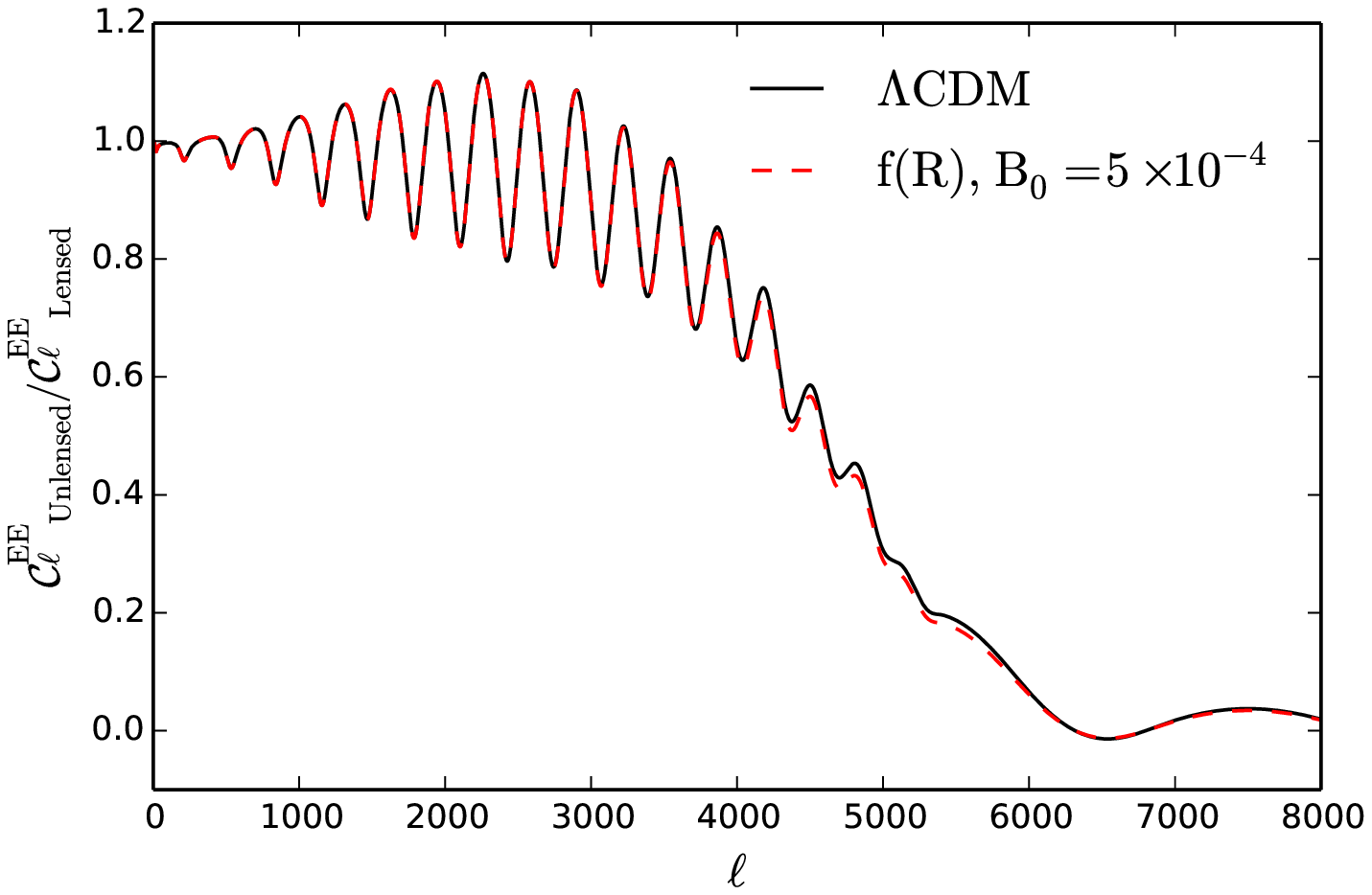} 
\caption{CMB lensing smears out the acoustic peaks, causing oscillatory 
enhancements and dilutions of the unlensed, primordial power. However, 
it also converts the higher amplitude $T$ modes into $E$-mode polarization, 
in particular on the characteristic scattering scale of $\sim2'$. 
This leads to almost all the $E$-mode power at $\ell>5000$ being due to 
CMB lensing. The black curve shows the ratio of unlensed to lensed 
$E$-mode power as a function of multipole for a \lcdm\ cosmology, while the 
dashed, red curve shows that the same behavior holds in $f(R)$ gravity. 
} 
\label{fig:clee} 
\end{figure}

However, we highlight in Fig.~\ref{fig:clee} a very interesting property 
of the $E$-mode spectrum: at high multipoles it is almost wholly due to 
lensing. This holds independent of the details of the physics modifications 
we discussed in this paper. As an example, Fig.~\ref{fig:clee} demonstrates 
the same conclusion in the case of the modified gravity models as well. 
High resolution CMB polarization 
experiments with $\sim1$ arcmin beams can reach $\ell=10000$ and our 
current knowledge indicates that the $E$-mode polarization signals should not 
be overwhelmed by foreground polarization, so this could be a promising 
avenue for exploration.

\section{Conclusions} \label{sec:concl} 

CMB lensing is a unique probe of the physics driving both expansion and 
growth over vast ranges of cosmic history, sensitive to $z\approx5$ and 
beyond. Experimental measurements are approaching percent precision in 
dozens of multipole bins, and now- or imminently-operating high resolution 
CMB polarization experiments such as ACTpol, POLARBEAR/Simons Array, and 
SPT-3G, and the next generation CMB-S4, 
can achieve this over large areas of sky. This opens windows on 
physics beyond that mapped by the temperature power spectrum, or unlensed 
polarization. Here we 
focused on physics that introduces a new scale, such as modified gravity, 
massive neutrinos, and cold dark energy. 

We explored the signatures of this physics, first through analytic 
approximations to build intuition on the effects, and then through 
rigorous numerical calculations, for example using a modified gravity 
Boltzmann code and nonlinear prescription calibrated by simulations. 
The analytic intuition works well at predicting the areas and qualitative 
behavior of deviation from \lcdm\ in the CMB lensing deflection power 
spectrum. 

Moreover, the shapes (angular dependence) of the deviations are fairly 
distinct between the various scale dependent physics origins. Sufficiently 
accurate measurements thus have the promise of distinguishing the nature 
of the physics behind detected deviations -- i.e.\ measuring the 
gravitational coupling $G_{\rm eff}$, the sum of neutrino masses 
$\sum m_\nu$, or the dark energy sound speed $c_s$. While hints of scale 
dependent variation are seen in current data, they are not yet statistically 
significant. 

For modified gravity, we found that $f(R)$ gravity with $f_{R0}\approx10^{-4}$ 
could give a deviation roughly the same in amplitude as a shift in the 
matter density of $\Delta\om=0.01$. Due to the projection of wavemodes, 
it is crucial by $\ell\approx200$ already to treat the nonlinear 
wavenumbers consistently, which we did using MGhalofit. 
Other recent applications of CMB lensing to test gravity (not necessarily 
scale dependence, as we focus on here) include \cite{14124454,150206599}. 
Neutrino masses, even at the level of $\sum m_\nu=0.05\,$eV, also give a 
couple of percent signal, 
with a characteristic shape. Cold dark energy only distinguishes itself 
from quintessence at low multipoles, where cosmic variance dominates. 

Neglecting to account for scale dependent physics despite its presence 
will generally bias other cosmological parameter estimation. This can 
also be an issue for delensing of $B$-modes: if the poorly reconstructed 
parts of the deflection spectrum (and their contributions to the $B$-mode 
lensing polarization) are employed assuming some model that 
lacks existing scale dependence, this can bias the fit of the tensor to 
scalar ratio $r$ of inflation. 

The scale dependent physics will also show up in other power spectra. 
For the galaxy power spectrum this may be difficult to separate from 
scale dependent galaxy bias, small scale nonlinearities, and baryonic 
effects. Other CMB power spectra have the lensing contributions mixed 
with the primordial spectra. However, with the advance of high resolution 
CMB polarization experiments capable of reaching $\ell=10000$, and the 
hint that polarized foregrounds are small enough for such arcminute 
scale data to be useful, it is exciting to note that $E$-mode polarization 
at $\ell\gtrsim5000$ is almost purely due to CMB lensing. CMB lensing, 
and its $E$- and $B$-mode contributions, is an arena capable of offering 
new physics insights beyond the concordance model.

\acknowledgments 

We thank Antony Lewis and Blake Sherwin for helpful discussions. 
AH is supported by an NSERC postdoctoral fellowship. 
EL is supported by the U.S.\ Department of 
Energy, Office of Science, Office of High Energy Physics, under 
Award DE-SC-0007867 and contract no.\ DE-AC02-05CH11231.


\end{document}